\documentstyle[12pt]{article}
\def\demi{{\textstyle {1\over2}}}

\def\bea{\begin{eqnarray}}
\def\eea{\end{eqnarray}}

\def\be{\begin{equation}}
\def\ee{\end{equation}}

\let\nonu=\nonumber

\begin{document}
\bibliographystyle{perso}

\begin{titlepage}
\null \vskip -0.6cm
\hfill PAR--LPTHE 02--39

%\hfill RUNHETC-2000-54

\hfill hep-th/0207184

\vskip 1.4truecm
\begin{center}
\obeylines

        {\Large	Going down from a 3-form in  16 dimensions}
\vskip 6mm
Laurent Baulieu
%$^a$
%, Alessandro Tanzini$^{a,b}$
{
%\em $^a$
 Laboratoire de Physique Th\'eorique et Hautes Energies,
 Universit\'es Pierre et Marie Curie, Paris~VI
et Denis~Diderot,~Paris~VII}
%{\em $^b$ Istituto Nazionale di Fisica Nucleare, Roma}
 %{\em and}
%{\em
% $^{\dag}$  Dept. of Physics, Rutgers University, New Brunswick,
%NJ~60637,~USA }

\end{center}

\vskip 13mm

\noindent{\bf Abstract}: Group theory indicates  the
existence of a $SO(8)\times SO(7)\subset~SO(16)$ invariant self-duality
equation   for a 3-form in 16 dimensions. It is  a    
  signal for   interesting topological field theories,
especially   on 8-dimensional manifolds with
 holonomy group smaller than or equal to  $Spin(7)$, with a gauge symmetry
that is
$SO(8)$ or
$SO(7)$. Dimensional reduction also provides new supersymmetric theories
in 4 and lower dimensions, as well as  a model with gravitational
interactions in 8 dimensions, which  relies on the octonionic  
gravitational self-duality equation. 
\vfill

\begin{center}

\hrule \medskip
\obeylines
Postal address: %
Laboratoire de Physique Th\'eorique et des Hautes Energies,
 Unit\'e Mixte de Recherche CNRS 7589,
 Universit\'e Pierre et Marie Curie, bo\^\i te postale 126.
4, place Jussieu, F--75252 PARIS Cedex 05

\end{center}
\end{titlepage}

\def\TQFT{ Topological Field Theory}
\section{Introduction}

  Self-duality equations  play an important role in 
the context of topological field theory (TQFT), by  providing topological
gauge functions that one enforces in a BRST invariant way. This often 
determines   supersymmetric actions in a twisted form.   There is a
classification  in~\cite{laroche} of    possible self-duality equations
for the curvatures of forms of degree
$p$  in spaces with dimension $d$, $^*G_{p+1} = T\wedge G_{p+1}$. Here,
$T$ is a 
 tensor invariant under a maximal sub-group of $SO(d)$ and 
$G_{p+1}$ is the curvature of the p-form. 
A requirement, for  determining a TQFT for a   form of a given degree, 
is that the number of self-duality equations for the curvature must
equal the number of degrees of freedom of the form, modulo its gauge
invariance (that is, the number of possible gauge covariant equations of
motion for the form). This gives, case by case, and depending on the
value of the space   dimension, certain conditions for  
$T$, which were  solved in~\cite{laroche},   by  using  the available
numerical Lie algebra tables. This numerical approach has limits the
analysis to spaces with dimensions lower than 16, for forms of degrees
less than 8.  A certain number of non-trivial possibilities were  
found. They are listed in the table page 11 of reference~\cite{laroche}.
This classification shows the existence   of cases that   go  beyond the 
obvious  self-duality equation   $G_n=^* G_n$ 
 of forms of degree~$n-1$ in dimension
$2n$, for which 
    T is the  $SO(2n)$ invariant 
antisymmetric tensor. 

This table determines for instance    the octonionic self-duality
equation~\cite{corrigan} for a Yang--Mills field in eight 
dimensions. The latter allows one to build  the 8-dimensional Yang--Mills
TQFT~\cite{bakasi}\cite{acharya}, which is  $SO(8)$ covariant  and
$Spin(7)$ invariant. It is a twisted version of
the 8-dimensional  supersymmetric Yang--Mills theory and it  
exhibits  a rich structure. By dimensional reduction and  a
suitable gauge-fixing in the Cartan algebra that is  allowed by
topological invariance, it gives  the abelian monopole theory of Seiberg
and Witten~\cite{swm}.  Other links with
$N=4,D=4$ models have also been exhibited in~\cite{bakasi}, as well  as
links to matrix models.

The aim of this paper is to study the implications of another prediction 
 of~\cite{laroche}, namely,   the existence of a self-duality equation
for the curvature $G_4$ of a 3-form
$B_3$ in 16 dimensions. This equation  is invariant under a maximal
subalgebra
$SO(8)\times SO(7)$ of $SO(16)$.
%~\footnote{ In comparison, the existing
%few   possibilities for other type of forms in other dimensions in the
%table imply a breaking of the Lorentz symmetry into symmetries that
%appear to us much less attractive.}.  
   We will be interested in  constructing  8-dimensional TQFT's that
possibly  descend  from this self-duality equation, by   
  dimensional reduction from 16
   to 8 dimensions. They depend on     fields coupled to the
genuine 
  8-dimensional Yang--Mills theory~\cite{bakasi},  but, now,  the  
gauge symmetry group is determined. These  fields  
descend from the 3-form, and their degrees of freedom can be spanned in
suitable representations of
$SO(8)$, or, possibly, of subgroups of $SO(8)$.   We
will directly build the curvatures in  8 dimensions. In  TQFT's, the
relationship between curvatures and forms is only    restricted by the
necessity of   Bianchi identities. As a consequence  of this freedom,
  one can get  interactions in lower dimensions, although one starts
from a free abelian 3-form in 16~dimensions. The determination of
possible curvatures reduces to a rather easy algebraic problem, as
indicated in earlier papers.  Our procedure   singles
out two possible gauge symmetries, either
$SO(8)$ or 
$SO(7)$~\footnote{More precisely, we should say
$SO(7)^\pm$, depending on the way $SO(7)$ is embedded in
$SO(8)$.}. These gauge groups are quite  relevant   for the
determination of instanton solutions. Indeed,  in  8
dimensions,  these  groups  play an analogous   
role    
to that of     $ SU(2)$  in   4 dimensions. The
t'Hooft symbols $\eta^a_{\mu\nu}$, which mix the space and internal
symmetry indices, and  express the instanton solutions in four dimensions,
are  replaced in eight dimensions by other symbols, which are related to
the octonion structure coefficients, as shown in
\cite{fubini}.  $G_2$ is also an interesting possibility for relabeling
the internal indices. By a further dimensional reduction from 8 to 4
dimensions,     new couplings to matter can be found.  By going down to
two dimensions we suggest  a connection with a Matrix theory description
of the Seiberg--Witten curves.

The dimensional reduction of the 16-dimensional model may also  have
an interesting gravitational interpretation.  We   suggest that the
degrees of freedom of the 8-dimensional theory  can be related to the
fields of twisted supergravity.  The vacua of this theory are related to
gravitational instantons with
$Spin(7)$ holonomy. Particular solutions of this kind have  recently been
studied in
\cite{floratos}\cite{kanno}. 

 Finally,
one of our  motivations is that a TQFT of a~3-form gauge field in higher
dimensions is a quite attractive candidate for generating the   $M$ or $F$
theory, as suggested by its central role in  $D=11$, $N=1$
supergravity.

\section{The theory in 16 dimensions}
\def\om{T}

Given a real 4-form in in 16 dimensions, \cite{laroche} indicates the
existence of an interesting  self-duality equation,
\bea \label{self-dualityc}
^*G_4=\om_8\wedge G_4,\eea
that is, $
G_{\mu\nu\rho\sigma }=
\om _{\mu\nu\rho\sigma\alpha\beta\gamma\delta} 
G_{\alpha\beta\gamma\delta}$, 
where  the fully antisymmetric $SO(16)$ self-dual tensor $\om_8$ is 
a singlet  under a maximal subalgebra   $SO(8)\times SO(7)  \subset
SO(15)\subset SO(16)$. The 4-form
$G_4$   can be decomposed  into a direct sum of terms that are  
irreducible under $SO(8)\times SO(7)$.  One of the
factors corresponds to  
a representation of dimension $455$. The   point is that
this number is precisely the number of components of a 3-form gauge field
in 16 dimension, defined modulo gauge transformations, $B_3\sim B_3
+d\Lambda$, since  $   (^{15}_{3}) =455$. This tells us that one can
interpret , (i),    
$G_4$ as  the curvature of a 3-form
$B_3$, and, (ii),   Eq.(\ref{self-dualityc})  
as a 16-dimensional
   self-duality equation, which is   $SO(8)\times
SO(7)$-invariant  and allows one  to determine 
$B_3$, modulo gauge
transformations. Moreover,
the
 decomposition of this representation under $SO(8)\times SO(7) $
is \cite{laroche} :
 \bea
\label{decomposition}
455=  (1,35)\oplus (8,21)\oplus (28,7)\oplus(56,1),  
\eea
This decomposition is suggestive  enough to indicate to  us the way  
   the various fields arising from the dimensional reduction
in 8 dimensions can be arranged in   group representations. 

In 16 dimensions, we can consider the following $SO(8)\times
SO(7)$-invariant topological term:
\bea \label{top}
{
\int _{M_{16} }  \om_8\wedge G_4\wedge G_4 
%+ G_4\wedge G_4\wedge G_4 \wedge G_4 
}
\eea
A BRST invariant gauge-fixing can be obtained  obtained    by  
adding the following $Q$-exact term, which gives a  16-dimensional
action: 
\bea 
\label{free}
\int_{M_{16}} d^{16} x 
\Big\{ Q, \chi^{\mu\nu\rho\sigma} 
\Big[(G_{\mu\nu\rho\sigma} -
\om_{\mu\nu\rho\sigma\alpha\beta\gamma\delta}\ 
G_{\alpha\beta\gamma\delta}) + 
\demi H _{\mu\nu\rho\sigma}\Big]\Big\}
\eea
Here $Q$ is a standard topological BRST operator for the 3-form $B_3$ and 
the antighost $\chi^{\mu\nu\rho\sigma} $ is   self-dual  in the sense of
Eq.~(\ref{self-dualityc}). We do not write the complete action, which
would necessitate the BRST invariant  gauge-fixing of the topological
ghosts  that occur in the definition of the topological BRST symmetry, as
detailed for instance in 
\cite{symtop}.

    Using the standard construction, Eq.~(\ref{free}) is expected  to 
  determines   an   action of the following form:
\bea 
\int _{M_{16} } d^{16} x \ (
G_{\mu\nu\rho\sigma }G^{\mu\nu\rho\sigma }
-\om_{\mu\nu\rho\sigma\alpha\beta\gamma\delta}\ 
G^{\mu\nu\rho\sigma }G^{\alpha\beta\gamma\delta}
 +{\rm supersymmetric \  terms})
\eea
In this  action,   $G_4=dB_3$.
%~\footnote{We could introduce
%16-dimensional gauge invariant interactions   by changing
%$G_4=dB_3$ into
%$G_4=dB_3+^*(dB_3\wedge dB_3\wedge dB_3).$}. 
 The  supersymmetric, i.e,  ghost dependent terms,       depend
on   the   tensor
$\om$.
 Thus the $SO(16)$ covariant action  is     only   
$SO(8)\times SO(7)$ invariant, as the topological term in
Eq.~(\ref{top}) depends on the  given expression for $\om_8$.

   We actually
do not intend to study a   16 dimensional theory. Rather, we  will
shortly  give attention to   its possible descendants in 8 dimensions,
which we will obtain  from  dimensional reduction arguments, and
    rearrangements of    degrees of freedom in relevant group
representations.  The triality that exists in 8 dimensions will be useful.

The projection in 8 dimensions is suggested by the invariance of $\om_8$.
At first sight, Eq.~(\ref{decomposition}) suggests
that the self-duality equation can be decomposed after reduction
in  8 dimensions into self-duality equations for the curvatures
of one 
3-form, eight  2-forms,  twenty-eight 1-forms, and fifty-six 
0-forms. The above mentioned rearrangement means for instance that   the
latter fifty-six equations 
will    be assembled into  7 Dirac like matrix equations, which  mix the
curvatures of these 56  scalar fields fields. Thus, we will use the
possibility of identifying the 56 scalars as 7 spinors of $SO(8)$.

 The counting is such that the number of all the projected 
self-duality  equations  is exactly the number of gauge invariant degrees
of freedom of the  forms. Indeed, the number of gauge invariant 
self-duality equation for the curvatures of 3-forms, 2-forms, 1-forms and
0-forms in eight dimensions are  35, 21, 7  and 1 respectively. 
Eventually,   such    self-duality
equations  can be enforced  through a BRST invariant TQFT, where  all
propagators are  fixed, and   the
gauge symmetries of forms are     encoded in an equivariant way. As
for   
  Lorentz invariance in eight dimensions, \cite{bakasi}
suggests that it will be at  least reduced to  
$ Spin(7)\subset SO(8)$,   prior to  an   untwisting that 
could  be allowed by triality. Let us recall that   for  the  genuine
Yang--Mills 8-dimensional TQFT,   the full
$SO(8)$ invariance is recovered after untwisting \cite{bakasi}. 

  In the next section, we will  detail  these points and    write  
self-duality equations that seem relevant to us in  8 dimensions.

\section{Interacting gauged TQFTs in 8 dimensions}

In the most naive approach, the  fields  that 
occur after  the dimensional reduction of the  abelian  3-form from 16 to
8 dimensions are   classified  in  antisymmetric
representations of an internal global 
$SO(8)$ symmetry.  
\bea
\label{naive}
B_{\mu\nu\rho} \ \ {\small{(\rm in } \   D=16)}\ \to (B_{\mu\nu\rho},
B_{\mu\nu}^a, A_{\mu }^{[ab]}, \Phi_{ }^{[abc]}  \ )\ \ {{(\rm in } \  
D=8})
\eea
 The representations in which the 8-dimensional fields   in the
right hand side of Eq.~(\ref{naive}) take their values, are of dimensions
1, 8, 28 and 56 respectively. The upper latin indices
$a,b,c,\ldots$ denote the  internal $SO(8)$ indices.

We  are free to change the interpretation of these indices, as one
often does in a
topological field theory, a  possibility  that we    understand as
the essence of a  twist. Moreover,   we can  gauge  the
internal 
$SO(8)$ symmetry by suitable redefinitions of the relation between
the forms and the curvature.  These redefinitions are constrained by the 
  necessity  of Bianchi identities for the curvatures. 

In 8 dimensions,   we can  identify   vector indices as
spinor indices.  We can interpret the  eight 2-forms
$B_{\mu\nu}^a$ as the  components of  a commuting
spinorial 2-form  field $B_{\mu\nu}^\alpha$, and the fifty-six 0-forms 
 $\Phi_{ }^{[abc]} $   as the  components of  seven commuting 
$1/2$-spin field $\Phi_{ }^{\alpha (i)}$, $1~\leq~i\leq
~7$, that is,  
\bea
 B_{\mu\nu}^a \to B_{\mu\nu}^\alpha
\quad
\quad
 \Phi ^{[abc]} \to  {\Phi_{ }^{\alpha (i)}}
\label{trial}
\eea At this stage,   $A_{\mu}^{[ab]}$ is  a $SO(8)$-valued 
gauge field. (We will shortly discuss a possible
modification of this interpretation.) The spinorial
index
$\alpha$ in Eq.(\ref{trial})   runs from 1 to 8.  The internal
index $i$  runs from 1 to 7  and can be interpreted as the index of a
fundamental representation of dimension 7 of a given group, for instance
$SO(7)$ or $G_2$.

In view of the possible gauging  of the 
internal covariance of $\Phi^{\alpha (i)}$ denoted by the index $i$,
one can chose    a preferred direction in    $SO(8)$ and  
  enforce the associated $SO(7)$ gauge symmetry.
$A^{[ab]}_\mu$ can be further  decomposed in a an   $SO(7)$-valued  gauge
field  
$A^{[ij]}_\mu$ with  21 components   and    
  a  vector field  $T^{[i]}_\mu$, which is valued in the
fundamental representation of $SO(7)$.  Analogously,  we can    split the
eight 2-forms in 
$B_{mu\nu}^\alpha$  as  $B_{\mu\nu}^\alpha\sim (B_{\mu\nu}^i,
B_{\mu\nu})$. 

We thus have two possibilities,
\bea
\label{naive8}
B_{\mu\nu\rho} \ \ {\small{(\rm in } \   D=16)}\ \to (B_{\mu\nu\rho},
B_{\mu\nu}^\alpha, A_{\mu }^{[ab]}, \Phi_{ }^{i\alpha}  \ )\ \ {{(\rm in }
\   D=8})
\eea
\bea
\label{naive7}
B_{\mu\nu\rho} \ \ {\small{(\rm in } \   D=16)}\ \to (B_{\mu\nu\rho},
B_{\mu\nu}^i,B_{\mu\nu},  A_{\mu }^{[ij]}, T_\mu^i ,\Phi_{ }^{i\alpha}  \
)\
\ {{(\rm in } \   D=8})
\eea

Both  choices allows us  to write covariant self-dual equations 
for the fields and to build   consistent eight-dimensional TQFTs. It must
be noted that  
  $SO(7)$ is a natural gauge group for
eight-dimensional instantons. In fact, the eight-dimensional generalization
of the 't Hooft symbols mix the left-over
Lorentz symmetry  $SO(7)$  with an internal $SO(7)$ symmetry group 
\cite{fubini}.
A third possibility exists, which is to reduce the gauge symmetry down to
$G_2$. In this  case, 7 gauge fields among the $A^{[ij]}$ must be  
reinterpreted  as 56 bosonic degrees of freedom, which merely amounts to a
duplication of
$\Phi^{i\alpha}$, a scheme that we will not discuss here.

   Using the
covariant derivative
$D=d+A$ with respect to the
$SO(8)$ or 
$SO(7)$  gauge field
$A$, we now introduce  covariant interactions  by  defining 
appropriately    the relations between   forms and  curvatures. We take 
 the following definitions for the fields in 
Eq.(\ref{naive8}):
\bea\label{free8}
 G_{\mu\nu\rho\sigma}  & =&\partial _{[\mu}B_{\nu\rho
\sigma]} 
\cr
 G_{\mu\nu\rho }^\alpha & =&D _{[\mu}B_{\nu\rho
 ]}^\alpha\cr
%\eea
%\bea\label{interactions}
 F_{\mu\nu}^{[ab]} & =&\partial _{[\mu}A_{\nu] }^{[ab]}
+A_{[\mu }^{[ac]} A_{\nu ]c}^{b}
\nonu \\
S_{\mu }^{\alpha (i)} &=& \partial _{\mu }\Phi^{\alpha (i)}
+  A_\mu^{[ij]}\Phi^{\alpha (j)}
%\cr
%\bar K_{\mu }^{\alpha (i)}  &=& \partial _{\mu }\bar\Phi^{\alpha (i)}
%+  A_\mu^{[ij]}\bar\Phi^{\alpha (j)}
, 
\eea
For the fields in Eq.(\ref{naive7}), there is some flexibility, and we
have the possibility of curvatures with more interactions
($C(A)=Tr_{SO(7)} (AdA +{2\over 3} AAA)$ is the Chern--Simons form):
\bea\label{free7}
 G_{\mu\nu\rho\sigma}  & =&\partial _{[\mu}B_{\nu\rho]} +c_{ijk} 
 F_{[\mu\nu}^{[ij]} K_{\mu] }  ^k
\cr
 G_{\mu\nu\rho }^\alpha & =&
(G_{\mu\nu\rho }^i=D _{[\mu}B_{\nu\rho
 ]}^i +  F_{[\mu\nu}^{[ij]} T_{\rho]j}\ ,\ 
G_{\mu\nu\rho }=
\partial _{[\mu}B_{\nu\rho
]} + C(A) _{\mu\nu\rho }  )
\cr
%\eea
%\bea\label{interactions}
 F_{\mu\nu}^{[ij]} & =&\partial _{[\mu}A_{\nu] }^{[ij]}
+A_{[\mu }^{[ik]} A_{\nu ]k}^{j}
\nonu \\
 K_{\mu\nu}^{i} & =&\partial _{[\mu}T_{\nu] }^{i}
+A_{[\mu }^{[ij]} T_{\nu ]j}^{} +B_{\mu\nu}^i
\nonu \\
S_{\mu }^{\alpha (i)} &=& \partial _{\mu }\Phi^{\alpha (i)}
+  A_\mu^{[ij]}\Phi^{\alpha (j)}
%\cr
%\bar K_{\mu }^{\alpha (i)}  &=& \partial _{\mu }\bar\Phi^{\alpha (i)}
%+  A_\mu^{[ij]}\bar\Phi^{\alpha (j)}
\eea

In both cases, the   curvatures have been constructed from the
requirement of fullfilling   Bianchi identities, which are easy to check. The TQFTs
that involve these curvatures must be defined in an 8-dimensional space
with
 holonomy group smaller or equal to $Spin(7)$, in order to enable a 
self-duality equation for the Yang--Mills curvature. 
%The difficulty that must be solved is to get a self
%duality equation for the curvatures of the 2-forms in 8 dimensions.
%
%For writing self-duality equations whose number
%fits  well with the number of gauge independent degrees of freedom of the
%scalars and   two-forms, we will use     contractions of the curvatures by
%gamma-matrices, thanks to our rearrangement of these fields   in
%spinor representations.
%

The 8-dimensional self-duality equations  that we choose for the fields in
Eq.(\ref{free8}) are :
\bea\label{self-dual1}
 G_{\mu\nu\rho\sigma} -\epsilon  _{\mu\nu\rho\sigma 
\mu'\nu'\rho'\sigma'} 
G^{\mu'\nu'\rho'\sigma'}& =& 0
\nonu \\
\epsilon  _{\mu\nu\rho\sigma \tau
abc}  (\gamma ^a  \gamma ^b \gamma ^c) ^\beta_\alpha
 G _{ \rho\sigma\tau
}^\alpha & =& 0
\nonu \\
 F_{\mu\nu}^{[ab]} -\Omega _{\mu\nu\rho\sigma}F^{\mu\nu [ab]} & =& 0
\nonu \\
\gamma ^\mu S_{\mu }^{\alpha (i)} &=& 0
%\quad
%\gamma ^\mu \bar S_{\mu }^{\alpha (i)} = 0
\eea
 For
the case of the field decomposition in  Eq.(\ref{free7}),we have a more refined
possibility, where $f^i_{jk}$ stand  for the structure coefficients of
$SO(7)$:
\bea\label{self-dual1}
 G_{\mu\nu\rho\sigma} -\epsilon  _{\mu\nu\rho\sigma 
\mu'\nu'\rho'\sigma'} 
G^{\mu'\nu'\rho'\sigma'}& =& 0
\nonu \\
\epsilon  _{\mu\nu\rho\sigma \tau
abc}  (\gamma ^a  \gamma ^b \gamma ^c) ^\beta_\alpha
 G _{ \rho\sigma\tau
}^\alpha & =& 0
\nonu \\
 F_{\mu\nu}^{[ij]} -\Omega _{\mu\nu\rho\sigma}F^{\mu\nu [ij]} & =& 
\Phi^{  \alpha [i}
\Big (\gamma_{[\mu} \gamma_{\nu]}\Big)_{\alpha\beta}  \Phi^{  j]\alpha}
\nonu \\
K_{\mu\nu}^{[i]} -\Omega _{\mu\nu\rho\sigma}K^{\mu\nu  j} & =&f^i_{jk}
\Phi^{  \alpha j}
\Big (\gamma_{[\mu} \gamma_{\nu]}\Big)_{\alpha\beta}  \Phi^{  k\alpha}
\nonu \\
\gamma ^\mu S_{\mu }^{\alpha (i)} &=& 0,
%\gamma ^\mu \bar K_{\mu }^{\alpha (i)} = 0
\eea

The equations for the curvatures of the one-forms are the octonionic
equations used in \cite{bakasi},  where  $\Omega _{\mu\nu\rho\sigma}$ is
the self-dual
$  Spin(7)
\subset SO(8)$-invariant tensor.  $\Omega$ is defined from the octonionic
structure coefficients  
 $c_{ijk}$, 
with  $\Omega_{8ijk}= c_{ijk}$.  It  allows one to irreducibly decompose
in a
$  Spin(7)  $-invariant way  
the representation $\underbar{28}$ of $SO(8)$ as the sum of  the
representations
$\underbar{21}$ and 
 $\underbar{7}$ of $SO(7)$. This explains why, as needed in 8 dimensions,
 the self-duality equations of the Yang--Mills curvature
 only count for seven independent
equations.

 The
$ (\gamma ^\rho) ^\beta_\alpha$ are the
$8\times 8$ eight-dimensional gamma matrices. Having arranged fields 
 in spinorial
representations is the key for  having  
  self-duality equations, which are first order equations. The existence of
$\Omega$     follows from that of a 
covariantly constant  spinor
$\eta$, with  $\Omega _{\mu\nu\rho\sigma}=^\dagger\eta (
\gamma^\mu\gamma^\nu\gamma^\rho\gamma^\sigma )\eta$, which 
gives a reparametrization invariant definition  of the closed 4-form
$\Omega _4$.   The spinor $\eta$ exists when    the space   has a
holonomy group
 $H\subset Spin(7)$. 

The   condition on $G_4$ is the obvious  $SO(8)$ invariant self-duality
condition for the abelian curvature of a 3-form in 8 dimensions. By
enforcing  this condition in a BRST invariant way, one gets a
   TQFT action, as explained in~\cite{bakasi}. 

The
second equation for the curvature  $G_3^\alpha$ of the  spinorial 
field  $B_2^\alpha$ is  
$SO(8)$-invariant, and  deserves  more explanation.
It is analogous to a Rarita--Schwinger equation, but   it involves
a 2-form spinor, instead of the gravitino, which is one-form spinor. This
equation counts as many conditions  as there are degrees of freedom in
$B_2^\alpha$, modulo gauge transformations 
$B_2^\alpha\to B_2^\alpha +D\Lambda^\alpha $, since it   is a
two-form that linearly  depends on $G_3$. The rest of the degrees of
freedom in $ B_2$ must be gauge-fixed, using the
techniques of equivariant gauge-fixing\footnote{In the case of the system
of Eq.(\ref{free8}), one must use the Batalin--Vilkoviski formalism, due
to the non closure of the gauge transformation for a charged 2-form, and
antifields are needed. In the case of the system of Eq.(\ref{free7}),
owing to the presence of the field $T\mu^i$, the standard BRST
technology is sufficient for completing the gauge-fixing. }, in a way
that    generalizes    the  completion of the gauge
fixing of
$A$, once   the seven gauge covariant conditions  $ F_{\mu\nu}^{[ab]}
-\Omega _{\mu\nu\rho\sigma}F^{\mu\nu [ab]} = 0$ have been inposed. 
Actually, the completion of the gauge-fixing  of the 2-form 
$B_2^\alpha $   is   inspired from that
  for a gravitino. It is:
\bea\label{gf}
(\gamma^\mu D_\mu) (\gamma^\nu B_{\nu\rho}) =0
\eea
It is then a simple exercise to show that the square 
$|\epsilon  _{\mu\nu\rho\sigma \tau
abc}  (\gamma ^a  \gamma ^b \gamma ^c)  
 G _{ \rho\sigma\tau
}|^2  $ is essentially  equal to
  $|G_{\mu\nu\rho\sigma}|^2
+
|\gamma ^\rho G _{\mu\nu\rho  }|^2
+|\gamma ^\nu\gamma ^\rho G _{\mu\nu\rho } |^2 $  plus a Feynman type
gauge fixing for the 2-form gauge field
$B_2$, when Eq.(\ref{gf}) is enforced. The derivation is however
lengthy and will be  explained elsewhere. The important point is that one
gets a $SO(8)$-covariant propagator for the 2-form.

The   other conditions on $A$ and $\Phi$ give   gauge interactions
that are of interest, thanks to the   couplings 
introduced  in the definitions of the curvatures.

Using suitable Lagrange multipliers and antighosts, one can write a
BRST-exact TQFT action, whose bosonic part is essentially the sum of
the squares of these four conditions, that is,
\bea\label{toy}
 \int d^8x (  &|G_{\mu\nu\rho\sigma}|^2
+
 G _{\mu\nu\rho }^\alpha
  G _{\mu\nu\sigma }^\alpha
+
 G _{\mu\nu\rho }^\alpha
(\gamma ^\rho
 \gamma ^\sigma) ^\alpha_\gamma G _{\mu\nu\sigma }^\gamma
 \nonu \\&
 +
 G _{\mu\nu\rho }^\alpha
(\gamma ^\nu\gamma ^\rho
 \gamma ^\tau\gamma ^\sigma) _{\alpha_\beta} G _{\mu\tau\sigma }^\beta
+ | F_{\mu\nu}^{[ab]} |^2 +
 | S_{\mu }^{\alpha (i)}  |^2
\nonu \\&+ \ {\rm boundary\  and\  supersymmetric\  terms} \nonu \\&+  \
{\rm  \ ordinary \ 
 gauge\ fixing\ terms}
\ ) 
\eea
Here we used the basic properties of the octonionic  4-form ~$\Omega$
\cite{bakasi}, 
$| F_{\mu\nu}^{[ij]}+\Omega_{\mu\nu\rho\sigma}F^{\mu\nu [ij]}|^2
=3| F_{\mu\nu}^{[ij]} |^2+{\rm boundary\ terms}$, a well as  
 gamma matrix identities. The supersymmetric terms involve higher
ghost interactions that are clumsy, but straightforward to derive.

We are in the presence of a very specific theory, which involves a charged
2-form,  in a new and 
interesting way, with gauge interactions. By construction, it possesses a
$Q$-symmetry.  It is possible that,   by using   triality, this
theory could be untwisted   into theory which is invariant under the 
Poincar\'e supersymmetry, or, perhaps, only  under   part of it.  This
question will not be studied here.
We wish to emphasize that,  when one uses  the field decomposition 
of Eq.(\ref{free7}),  the    SO(7) gauge symmetry of the 
non-abelian   two-form follows from  our definition of    
curvatures  
$G_3=DB+FT$, $K=B_2+DT$ with  Bianchi identities $DG_3= FK_2$,
$DT_2=G_3$,  as in  \cite{bath}. 
 Eventually, this allows one to complete the gauge-fixing of ordinary
gauge symmetries   in the standard BRST way, without having to use the
 Batalin--Vilkoviski formalism,  (one has in this case a first rank BV
system), contrarily to  the case of the
 decomposition of  Eq.(\ref{free8}).
 Moreover, we       see that  the right-hand sides  of the
self-duality equations   in  Eq.(\ref{self-dual1}) involve   terms  that  
are similar to those used in the four-dimensional TQFT  for  monopoles
\cite{swm}, which also derive from dimensional reduction of a  theory in
higher dimension   (in this case an eight-dimensional one \cite{bakasi}). 
Finally, the presence of Chern--Simons   terms in the curvatures of $B_2$
and
$B_3$  give    gauge symmetries that     mix  
the gauge transformations of forms with   Yang--Mills gauge
transformation. If, in the process of dimensional reduction, one has
the 
creation of an anomaly, these modifications of the curvatures and of the 
gauge transformations might be  of interest for their cancellation.

\subsection{ A supergravity interpretation}

We now turn to another suggestive  interpretation of the fields
$A^{ab}_\mu$ and
$\Phi^{abc}$.  We can   interpret the twenty-eight one-forms $A^{ab}_\mu$
as a spin connection for the 8-dimensional manifold, $\omega^{ab}=B^{ab}$,
and the fifty-six zero-forms $\Phi^{abc}$ as the components of a
constrained vielbein $e^{a}_\mu$, which is appropriate for an
8-dimensional manifold with $Spin(7)$ holonomy, such that,  $e^{8}_8=
1$, $e^{i}_7=e ^{7}_i=\phi^i$ are described by 
 7 linear combinations of
the 56 fields $B^{abc}$'s and 
 $e^{i}_j$, $1\leq i,j~\leq 7$ are described by  
 49   other  independent combinations of the  $B^{abc}$'s, with:
\bea\label{achtbein1} e^{a}_\mu=\pmatrix {1&\phi^i \cr^t\phi^i &e^{i}_j}
\eea
%If we exhaust the topological freedom of $\phi^i$, we can put the
%vielbein under the following form, which is suitable  for manifolds with
%$Spin(7)$ holonomy:
%\bea\label{achtbein} e^{a}_\mu=\pmatrix {1&0 \cr 0 &e^{i}_j}
%\eea 
The decomposition of the 16-dimensional  3-form after reduction  
  to eight  dimensions is now of a purely gravitational nature : \bea (
B_{\mu\nu\rho} , B_{\mu\nu}^\alpha , \omega_\mu ^{ab}, e^{a}_\mu ) \eea
The interpretations of the topological ghosts and antighosts for
$e^{a}_\mu $ are as  the twisted   gravitino of $N=1$ supergravity
in eight dimensions, adapted to the vielbein in Eq.(\ref{achtbein1}).
%
%We now define the following Cartan curvatures for $e$ and $\omega$:
%\bea
%T_{\mu\nu} ^{a} =\partial _{[\mu}e_{\nu ]}^{a}
%+\omega_{b[\mu} ^{a}e_{\nu ]}^{b}
%,
%\quad
%{ R}_{\mu\nu} ^{ab} =\partial _{[\mu} \omega_{\nu ]}^{ab}
%+\omega_{c[\mu} ^{a}\omega_{\nu ]}^{cb}
%%+\lambda  e_{[\mu} ^{[a}e_{\nu ]}^{b]} \ \ .
%.\label{r}
%\eea
%%where the last term in (\ref{r})
%%Here, $ \lambda $ is a real parameter.

For    gauge-fixing the topological
freedom  for $ \omega_\mu ^{ab}$, we can  choose the torsion free
condition  $
T_{\mu\nu} ^{a} =0 $, which allows one to eliminate $\omega$ as a
function of $e$, in the standard way.
 
The  manifold has  $Spin(7)$ holonomy,  in such a way that it contains 
the invariant closed 4-form $\Omega_4$. In this case, the gravitational
instanton equation is just 
 \bea \omega ^{ab}_\mu= \Omega ^{abcd}\omega ^{cd}_\mu,
\ \rm that\ is, \ \omega ^{ab-}_\mu=0,\eea
 which counts as 7x8=56
independent equations. 
It is   relevant to use these 56 independent
equations as the topological gauge functions for exhausting the topological
gauge freedom in the vielbein $e^a_\mu$  in
Eq.(\ref{achtbein1}).   The construction of topological gravity in 8
dimensions has been  recently presented in   \cite{bbt}. 

Since we  predict  an action with a propagating metric, 
  the 3-form gauge field $B_{\mu\nu\rho}$ and the spinorial fields
$B_{\mu\nu}^\alpha$ are now subject to gravitational interactions by mean
of a BRST exact action as in Eq.(\ref{toy}).   These TQFT's      are
likely    to describe   invariants, which    are related
to the existence of gravitational instantons \cite{floratos}.
%; Moreover,
%they    can be coupled to the topological Yang--Mills theory of
%\cite{bakasi}.

\section{Dimensional reduction into renormalizable theories in 4
dimensions and below}
We now consider  further dimensional reductions,   
down to  4 dimensions. One motivation is of obtaining 
     new renormalizable  models, that contain  
  abelian  monopoles, with 
coupling to 
  supersymmetric matter. These models are in the spirit of the work of
Seiberg and Witten, for getting effective theories that allow for
unambiguous computations of   microscopic theories that are purely non abelian.
They include the  models that Seiberg introduced. 

As explained in
\cite{bakasi},  by 
 dimensionally reducing in four
dimensions  the
8-dimensional  octonionic   Yang--Mills  equation
$^*F=\Omega\wedge F$, one obtains   the coupled non-abelian
equations :
\bea\label{ncsw}
F^+_{\mu\nu}
=   [\bar M, \Gamma_{\mu\nu} M ],
\quad
\Gamma^\mu D_\mu M =     0
\eea
Here, $M^\alpha$ is a  Weyl spinor whose two complex     commuting
components  are  made from the gauge field components $A_5,A_6,A_7,A_8$:
\bea
 M^1= A_5 +i A_6, \quad M^2= A_7 + i A_8.
\eea
Thus $M=M^{A} T^A$ is
valued in the same Lie algebra as $A=A^{A} T^A$. A further gauge fixing
of the fields 
 in the
Cartan algebra is allowed by the topological gauge invariance, 
and one recovers   from Eq.~(\ref{ncsw}) the abelian
Seiberg--Witten equation 
\bea\label{csw}
F^+_{\mu\nu}
=    \bar M \Gamma_{\mu\nu} M ,
\quad
\Gamma^\mu D_\mu M   =   0,
\eea
where $M$ has the interpretation of a monopole. $M$ and its topological
ghosts build  a chiral matter multiplet  after untwisting.
 Let us explain the mechanism when the gauge symmetry is
$SU(2)$. In this case,  the projection in the Cartan algebra
 means
$A_{\mu}^{(1)}=A_{\mu}^{(2)}=0$ and
$A_5^{(3)}=A_6^{(3)}=A_7^{(3)}=A_8^{(3)}=0$, 
%and $A_5^2=A_6^1=A_7^2=A_8^1=0$, 
where the upper  indices are $SU(2)$ indices.
The Seiberg-Witten monopole 
%commuting complex Weyl spinor 
with charge plus or minus one with respect to the abelian gauge field is
simply given by the linear combination 
$M^{\pm}={1\over{\sqrt{2}}}\Big( M^{(1)} \pm M^{(2)} \Big)$.
%to the is nothing but :
%\bea
% M={\pmatrix { A_5^1+iA_6^2\cr A_5^1+iA_6^2}}
%\nonu\\
%\eea

For our theory in 8 dimensions, if we choose the case
of a $SO(8)$ gauge group,
the Cartan algebra is made of 4 independent $U(1)$ symmetries. These
symmetries will act with certain charges on the 56 scalars, when one
decomposes these representation $\underline{56}$ of $SO(8)$ on the four
$U(1)$. It gives other charges when the 56 scalars are  assembled into
seven spinors. If we restrict our-self to the maximal projection on one
of the
$U(1)$ subalgebra of the
$U(1)^4$ Cartan algebra of $SO(8)$, only a certain number of the   56
scalars will remain  coupled to the remaining abelian gauge field.
It is interesting to observe that,  depending on the $U(1)$ that one
chooses, one gets the two possible values of the charge that are related
by electromagnetic duality.

%If we choose the case of a $SO(7)$ gauge
% theory in 8 dimensions, another analysis must be done. 

\section{To matrix models}

The strategy of obtaining exact solutions of ${\cal N}=2$ gauge theories
from embedding in higher dimensions naturally emerges also in the context 
of M-theory. One of the notable applications of this theory is indeed the
solution of four-dimensional ${\cal N}=2$ models via the analysis of suitable
M-fivebrane backgrounds \cite{mwitt}. On the other hand, an explicit 
formulation of the M-theory is conjectured to be given by a matrix-model
\cite{banks}. We recall that in \cite{nekrasov} was explored the 
possibility of formulating a covariant action for the matrix strings 
from dimensional reduction of eight-dimensional Topological Yang-Mills 
Theory. It is interesting to observe that the dimensional reduction of our
topological model on a two-torus together with a suitable gauge-fixing
which set to zero all the fields but four components of the gauge field,
say for example $(A_4, A_5, A_6, A_7)$, gives rise to the equations
\bea
F = {i \over 2} [ \phi , {\bar \phi} ],
\quad
{\bar {\cal D}} \phi = 0
\label{self}
\eea
where
${\cal D} = D_6 + i D_7$, $F = {i \over 2} [{\cal D}, {\bar {\cal D}} ]$,
$\phi = A_4 + i A_5$ and the gauge connection is $A = A_6 + i A_7$.
The same equations (\ref{self}) can be derived from toroidal compactification 
of the Matrix theory in a suitable five-brane background, see Eq.(2.2) of 
\cite{gukov}, 
and describe the exact vacua
of an ${\cal N}=2$ four-dimensional model with an adjoint hypermultiplet.
Moreover, it can be shown that the brane configuration of \cite{mwitt}
corresponding to this model represents the same brane background as the 
Matrix theory setup \cite{gukov}.

{\bf Acknowledgments}: We thank M. Bellon, B. Pioline   and A. Tanzini
for very  useful discussions concerning this work. 

%\newpage


\begin{thebibliography}{10}

\bibitem{laroche}
  L. Baulieu,    C.
Laroche,
 { \it On Generalized self-duality Equations Toward
Supersymmetric   Quantum Field Theories of Forms,  }     {  
  \newblock
 Mod.    Phys.    Lett.    {\bf  A13}   1115-1132,  1998},   
hep-th/9801014.  

 \bibitem {corrigan}
 E.~Corrigan, C.~Devchand, D.~B.~Fairlie, J.~Nuyts, {\it First 
Order Equations for Gauge Fields in Spaces of Dimension Greater Than Four},
   \newblock
Nucl. Phys. {\bf {B214}}
(1983)452.





\bibitem{bakasi}
L. Baulieu,   Hiroaki Kanno,  
I.   M.    Singer,  
 { \it Special Quantum Field Theories In Eight And Other
Dimensions,  }  
\newblock {     Commun.    Math.    Phys.       {\bf  194}   149, 
1998},   hep-th/9704167.  

\bibitem{acharya}
B.S.Acharya, M. O'Loughlin, B. Spence,
{it  Higher Dimensional Analogues of Donaldson-Witten Theory,}
 \newblock{
Nucl. Phys. {\bf {B503}}
(1997) 657}, hep-th/9705138.


\bibitem{swm}
 E. Witten,  
 {  \it Monopoles and Four-Manifolds,  
}
\newblock   {      Math.    Res.    Let.   
Phys.    {1}   (1994) 769},   hep-th/9411102.


\bibitem{fubini}
 S. Fubini,    H.
Nicolai,
  { \it The Octonionic Instanton,  }     {  
  \newblock
     Phys.    Lett.    {\bf  155B}    (1984) 369};  
 D.~B.~Fairlie, J.~Nuyts,
 {   }     {  
  \newblock
J.    Phys.        {\bf  A17}   (1984) 2867.}   



\bibitem{symtop}
 L. Baulieu,  
 {  \it On The Symmetries Of Topological Quantum Field Theories,  
}
\newblock   {      Int.    J.    Mod.   
Phys.    {\bf A10}   4483-4500,  1995},   hep-th/9504015.
 


\bibitem{bath}
 L. Baulieu, J. Thierry-Mieg,  { \it On a New Type of Gauge
Transformations}, 
{\newblock       Phys. Lett. {\bf 144B} 221,1984.}


\bibitem{floratos}
 E.G. Floratos, A. Kehagias
{\it Eight-Dimensional self-dual Spaces}
hep-th/9802107,  Phys. Lett. {\bf  B445} (1998) 69,
I. Bakas, E. Floratos, A. Kehagias,
{\it  Octonionic Gravitational Instantons }
 hep-th/9810042, Phys. Lett. {\bf  B427} (1998) 283.


\bibitem{kanno}
H.~Kanno and Y.~Yasui,
{\it On Spin(7) holonomy metric based on SU(3)/U(1)},
hep-th/0108226.

\bibitem{mwitt}
E.~Witten,
{\it Solutions of four-dimensional field theories via M-theory},
Nucl.\ Phys.\ B {\bf 500} (1997) 3
[hep-th/9703166].

\bibitem{banks}
T.~Banks, W.~Fischler, S.~H.~Shenker and L.~Susskind,
{\it M theory as a matrix model: A conjecture},
Phys.\ Rev.\ D {\bf 55} (1997) 5112
[hep-th/9610043].

\bibitem{nekrasov}
L.~Baulieu, C.~Laroche and N.~Nekrasov,
{\it Remarks on covariant matrix strings},
Phys.\ Lett.\ B {\bf 465} (1999) 119
[hep-th/9907099].


\bibitem{bata}
L.~Baulieu, A.~Tanzini,
{\it Topological 
Gravity versus Supergravity },
[hep-th/9907099].

\bibitem{bbt}
L.~Baulieu, B.~ bellon, A.~Tanzini, 
{\it  Eight-Dimensional Topological Gravity and its Correspondence with
Supergravity}, hep-th/0207020.

\bibitem{fre}
L.~Anselmi, P.~Fre,
{\it $N=2$ Supergravity as a twisted Topological 
Gravity},

\bibitem{gukov}
S.~Gukov,
{\it Seiberg-Witten solution from matrix theory},
hep-th/9709138.

\end{thebibliography}
\end{document}